\documentclass[aip,jcp,numerical,twocolumn, reprint]{revtex4-1}

\usepackage{graphicx}% Include figure files
\usepackage{url}
\usepackage[version=3]{mhchem}
\usepackage{siunitx}
\usepackage{amsmath,amssymb}
\usepackage{epsfig}
\usepackage[utf8]{inputenc}
\usepackage{threeparttable}
\pdfoutput=1

\begin{document}
\setlength{\parskip}{0.3\baselineskip}
%\preprint{}

\title{Crystal nucleation mechanism in melts of short polymer chains under quiescent conditions and under shear flow}

\author{Muhammad Anwar}
\author{Joshua T Berryman}
\author{Tanja Schilling} 
\affiliation{Universit\'e du Luxembourg, Theory of Soft Condensed Matter Physics, Physics and Materials Research Unit, L-1511 Luxembourg, Luxembourg}


\date{\today}

\begin{abstract}
We present a molecular dynamics simulation study of crystal nucleation from undercooled melts of n-alkanes, and we identify the molecular
mechanism of homogeneous crystal nucleation under quiescent conditions and under shear flow. We compare results for $n$-eicosane (C20) and $n$-pentacontahectane (C150), i.e.~one system below the entanglement length and one above.
Under quiescent conditions, we observe that entanglement does not have an effect on the nucleation mechanism. For both chain lengths, the chains first align and then straighten locally, then the local density increases and finally positional ordering sets in.  At low shear rates the nucleation mechanism is the same as under quiescent conditions, while at high shear rates the chains align and straighten at the same time. 
We report on the effects of shear rate and temperature on the nucleation rates and estimate the critical shear rates, beyond which the nucleation rates increase with the shear rate. In agreement with previous experimental observation and theoretical work, we find that the critical shear rate corresponds to a Weissenberg number of order 1. Finally, we show that the viscosity of the system is not affected by the crystalline nuclei.
\end{abstract}

%\keywords{}

\maketitle


When a liquid is cooled below its crystal-liquid coexistence temperature, 
crystallites are formed. The shapes, sizes and structures of these 
crystallites strongly influence the properties of the final, solidified 
material. This is particularly relevant for polymers, which generally do not 
reach a perfect single crystalline state but remain poly- or semicrystalline 
after cooling. 

Polymer melts often flow during processing. Flow can change crystal nucleation 
and growth processes and hence affect the materials 
properties of crystalline and semicrystalline plastics. 
Understanding crystallization in flowing polymer melts is thus a topic of 
technological relevance. But it is also a challenging topic from the point 
of view of basic theoretical physics, because relaxation in polymer melts 
occurs on a hierarchy of time-scales 
that spans several orders of magnitude. When discussing phase transitions 
in polymers, one inevitably deals with 
non-equilibrium processes, which can only  to a very limited extent be 
described by quasi-equilibrium approaches. This fact poses a serious
challenge to any attempt to theoretically model polymer crystallization.

In spite of intensive research efforts since the early 1940s, the molecular 
mechanism of polymer crystallization is still not completely understood\cite{Muthukumar2004}.
Experimental research has been carried out using a wide range of
techniques both, on polymers under quiescent conditions 
\cite{Imai1994,ImaiM1995,Ezquerra1996,Keller1994,ImaiM1992,Strobl2000a,Strobl2005,Strobl2006,G.Strobl2007,Strobl2009a}
 and in external fields \cite{Somani2002a,Somani2002,Abuzaina2002,Lellinger2003,Coppola2004,Elmoumni2006,Acierno2008,Zhang2012}. Crystallization rates and critical shear rates have been measured for different polymeric 
materials, the morphological features of the final crystal structure and the effect of molecular weight on the crystallization kinetics have been studied. 
But  the primary nucleation mechanism has not been identified, because the short length- and time-scales on which it takes place are difficult to access experimentally.

 Most theoretical approaches to flow induced crystallization are based on 
coarse-graining. Generally, sets of coupled differential equations for the 
time evolution of macroscopic quantities (such as e.g.~the volume occupied 
by crystallites or the thickness of lamellae) are derived partly from the 
underlying microscopic theories, partly from balance conditions, and from
considerations regarding
the structure of effective free energy landscapes (see e.g.~refs.~[\cite{Custodio2009, Coppola2004, vanMeerveld2008, Zuidema2001, M.Doi1986, Shimada1988,P.D.Olmsted1998, HonggeTan2003,Kaji2005}]). While undoubtedly useful, these
models are inevitably semi-empirical. Coarse-graining requires approximations 
already in the equilbrium case. For the non-equilibrium case, in which one 
usually does not know the probability distributions of microstates 
according to which 
state-space averages would need to be taken, no systematic approach exists.

As the molecular length- and time scales involved in nucleation and growth  
processes are below experimental resolution, and a theoretical approach 
is challenging because of the the full non-eqilibrium nature of the problem,
computer simulations are a promising alternative method to solve the problem.
McLeish, Olmsted and co-workers have over the past 15 years developed a 
comprehensive set of theoretical and computer simulation techniques and 
experimental model systems to study polymers under flow. To address 
crystallization they derived a kinetic Monte Carlo algorithm on the basis of 
kinetics extracted from the GLaMM model \cite{Graham2003}, embedded 
it in a Brownian dynamics simulation \cite{Graham2009, Graham2010} and extended this approach by a fast nucleation algorithm to compute nucleation 
rates \cite{Jolley2011a}. This model 
captures many features of flow induced crystallization, however,
parts of it are based on an effective free energy picture i.e.~on the 
assumption of separating relaxation time-scales and thus quasi-equilbrium.

Atomistic computer simulations have been used to study polymer crystallization
 under quiescent conditions
\cite{Esselink1994,Takeuchi1998,Fujiwara1998,Fujiwara1999,Yi2009,Yi2011,Yi2013,Zerze2013,Anwar2013,Yamamoto2013,Yamamoto2010,Yamamoto2008,Yamamoto2004,Yamamoto1998,Luo2011,Luo2009,Welch2001,Meyer2001,Sommer2000,Waheed2002,Muthukumar2005,Muthukumar2003a,Muthukumar2000}
and under flow or large deformation \cite{Koyama2002,Koyama2003,Lavine2003a,Ko2004a,Ionescu2006,Jabbarzadeh2009a,Dukovski2003,Baig2010,Baig2010a}.
Most of these studies focus on the growth process rather than the nucleation 
process, because
nucleation is by definition a rare event (an event that occurs on a 
time-scale much larger than the time-scale of the local dynamics) and 
therefore difficult to tackle by atomistic simulation.
Nucleation in short chain alkanes under quiescent conditions has nevertheless 
been simulated \cite{Esselink1994,Takeuchi1998,Fujiwara1998,Fujiwara1999,Yi2009,Yi2011,Zerze2013,Anwar2013} and a scenario for the nucleation mechanism has 
been identified. (We will refer to this mechanism in detail in the results section.)  
The first direct computation of homogeneous nucleation rates in long chain 
alkanes by means of computer simulation has recently been presented by 
Rutledge and co-workers\cite{Yi2013}.
Their work was focussed on the nucleation and growth rates and the free energy 
landscape associated with the crystallization process rather than the 
microscopic mechanisms. 

To our knowledge, there is no simulation study yet that resolves the 
molecular nucleation mechanism in polymers under shear.
In this article, we present a detailed analysis of the formation of 
crystal nuclei from the melt in short chain 
alkanes under shear and in long chain alkanes under quiescent and shear 
conditions.

\section{Model \& Order parameters}
We have used a united atom model for polyethylene that has been proposed by 
Paul et al.\cite{Paul1995} and later modified by Waheed et al\cite{Waheed2002}. 
(For a table of interaction parameters we refer the readers to our previous 
work \cite{Anwar2013}.)
In order to carry out the simulations by means of the ESPResSo package \cite{Limbach2006} we implemented (and have made available) the dihedral-cosine potential and Lees Edwards periodic boundary conditions, which were not previously supported by ESPResSo.

We used several order parameters to identify the crystallites in the melt:  for the analysis we split the long chains (C150) into segments of 15 monomers, while we regarded the short chains (C20) as single segments. Then we computed the radius of gyration $R_{g}$ of each segment and the nematic order parameter $S_2$ of those segments that were involved in the formation of the critical nucleus. (A 
definition and detailed description of these parameters can be found in our 
previous work on C20 \cite{Anwar2013}.) 
Further we measured the local alignment of bonds: Monomomers within a radius $r_c=1.4\sigma$ were considered as neighbours, where $\sigma$ is the length scale set by the Lennard Jones interaction in the polymer model. Two neighbours $i$ and $j$ were considered as ``aligned'' if the chains they belonged to locally were almost parallel $(\theta_{ij} \leq \ang{10})$. For a particle to be considered ``crystalline'', it had to have at least 13 aligned neighbours in case of C20 and 12 aligned neighbours in case of C150. These numbers were obtained by sampling the probability distributions of the number of aligned neighbours in the bulk crystal and the bulk liquid.

\section{C150 under quiescent conditions}
\subsection{Simulation details}

First we discuss the nucleation mechanism in a quiescent system of 
$n$-pentacontahectane(C150). We chose C150, because it has the 
minimum length for which we can capture the effects of 
entanglement on crystallization  and observe a folded 
chain crystal structure (the entanglement length has been 
reported to be between 60 and 
90 monomers\cite{Qin2011,Hoy2009,Subramanian2008,Tanaka2000}).
We simulated 100 chains at 280K, which corresponds to 30$\%$ supercooling. (For the model that we use, the equilibrium melting 
temperature of C150 is 396.4K\cite{Yi2013}.)
We equilibrated the system at 500K, i.e.~well above the melting 
temperature.
After equilibration we quenched the configurations from 500K to 280K and 
observed the nucleation event. 
We performed these simulations under constant pressure and constant 
temperature conditions. The pressure was fixed at 1 atmospheric pressure.

The polymer model contains a Lennard-Jones-type interaction term. 
We therefore use Lennard Jones units to present our data (i.e.~the particle 
mass $m$, the interaction energy $\epsilon$ and resulting timescale 
$\tau = \sqrt{m \sigma^2/k_{B}T}$). Quantities which 
can be compared directly with the experimental results are presented 
in SI units. We used a Langevin 
dynamics based thermostat and barostat\cite{Kolb1999}. 
The friction coefficient $\gamma$ used for the thermostat was 1.0$\tau^{-1}$ 
and the piston mass for the barostat was 0.00001$m$.

\subsection{Nucleus formation}
To determine the induction time and the size of the critical nucleus
we performed a mean first passage time analysis\cite{wedekind2007} on 20 
independent trajectories. (This 
method has been successfully applied to simulation data of nucleation in 
$n$-alkanes before \cite{Yi2009,Yi2011,Yi2013}.) 
The values for the induction time $t^{*}$ and the number of particles in the critical nucleus $n^{*}$ are given in Table~\ref{tab:table1}. We find the nucleation rates to be in rough agreement with the results of Yi et al.~\cite{Yi2013}. As we were using slightly different system sizes, different barostats and thermostats, small differences in the results were expected.

\begin{table}
\centering
\caption{Results of the mean first passage time analysis for C150 at 280 K.}
\label{tab:table1}
\begin{tabular}{|l|c|c|c|}
  \hline
Study & $n^*$ & $t^*$(ns) & $I(10^{25}cm^{-3}s^{-1})$ \\
\hline
Yi.et .al \cite{Yi2013} &  143$\pm$14 & 293$\pm$19 & 1.47$\pm$0.10  \\ 
This work &  87$\pm$9 & 354$\pm$41 & 0.72$\pm$0.08  \\ 
\hline
\end{tabular}
\end{table}

\begin{figure}[t]
\centering
\includegraphics[width=0.5\textwidth]{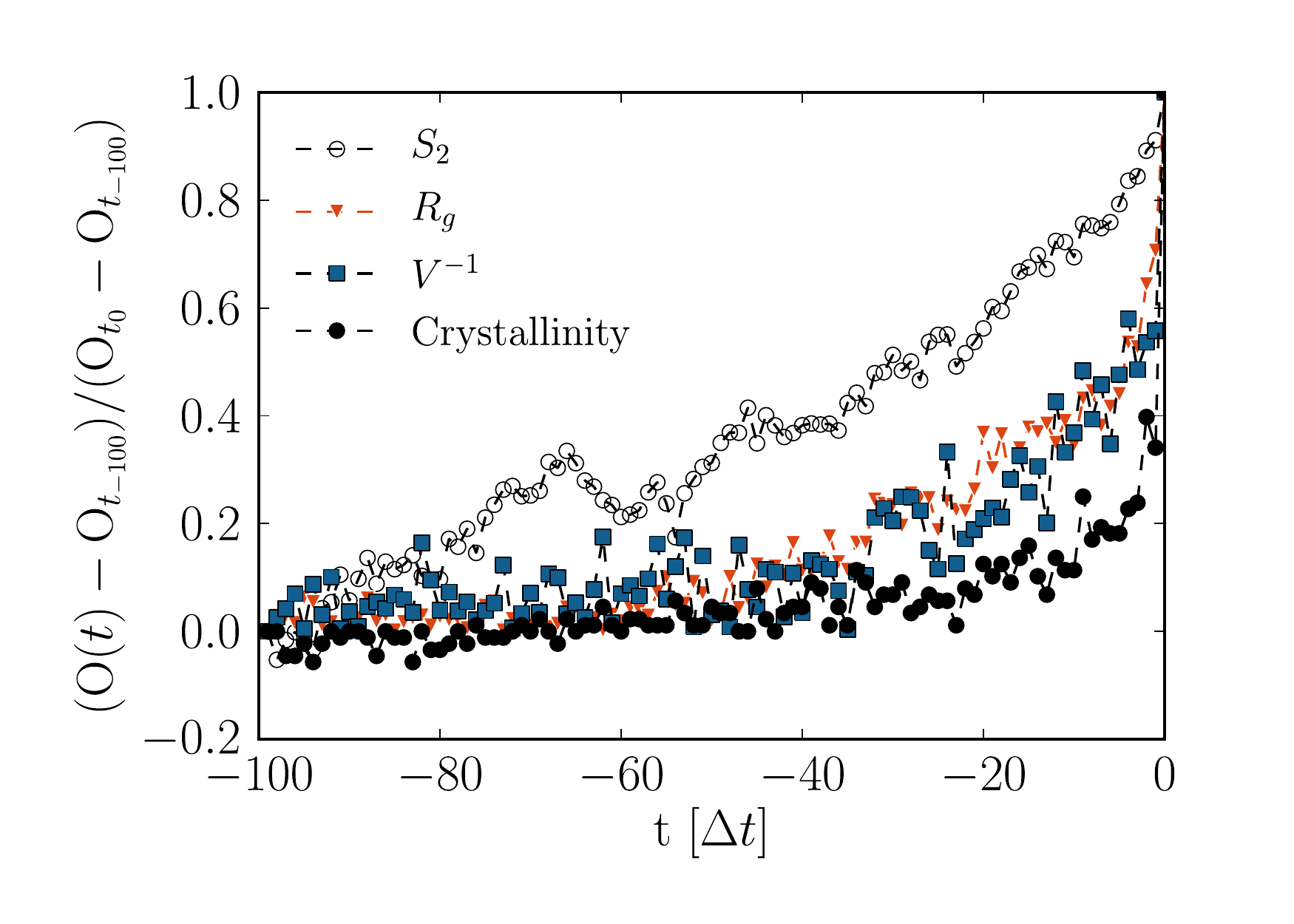}
\caption{Relative variation of several observables (O) from the melt to the formation of a critical nucleus, computed for those particles that are part of the nucleus at the nucleation time $t=t_{0}$: orientational order $S_{2}$ (black, open circles), radius of gyration $R_g$ (red, triangles), 
the inverse of the Voronoi cell volume $V$ (blue, squares) and the crystallinity order parameter (black, closed circles). 
The curves are averaged over 20 independent trajectories progressing backward 
in time from the nucleation time $t=t_{0}$ in steps $\Delta t=100000 \tau$ to $t=-100\Delta t$.} 
\label{fig:fig_1}
\end{figure}

To analyze the nucleation mechanism, we identify in each trajectory 
those particles 
that are part of the critical nucleus at the nucleation time $t_0$. We then
trace them backwards in time and compute their structural and orientational 
properties. We proceed backwards until the particles are 
indistinguishable from the melt particles. For 20 independent 
trajectories we compute the 
average radius of gyration $R_{g}$ of all chain segments that are part of the 
nucleus at $t_0$, the nematic order $S_2$ of 
these chain segments, the average volume $V$ of the 
Voronoi\citep{Rycroft2009} cell associated to each particle that is 
part of the nucleus and its crystallinity order parameter.
In Fig.~\ref{fig:fig_1} we show the relative variations of these quantities with respect to the values they had at $-100\Delta t$, where $\Delta t=100000 \tau$.
When we advance from the supercooled melt towards the formation of the critical nucleus at $t_0$, we observe first an increase in the global 
orientational order $S_2$, then an increase in the radius of gyration of the 
segments and in the local density, and finally the crystal structure is formed.
We conclude that the nucleation mechanism in long, 
entangled chains is the same as in short, non-entangled chains: orientational ordering precedes straightening \cite{Anwar2013}. 

Note that the Voronoi volume per particle in the nucleus does not deviate from its melt value until the very late stages of the nucleation process. We are thus not dealing with the spinodal decomposition assisted crystallization process that has been proposed by Olmsted \cite{P.D.Olmsted1998}.  Our results also stand in contrast to the scenario suggested by Doi et al.~in which crystallization is initiated by an increase 
in the persistence length, followed by the alignment of the chains 
\cite{M.Doi1986, Shimada1988}.

\begin{figure}[t]
\centering
\includegraphics[width=0.5\textwidth]{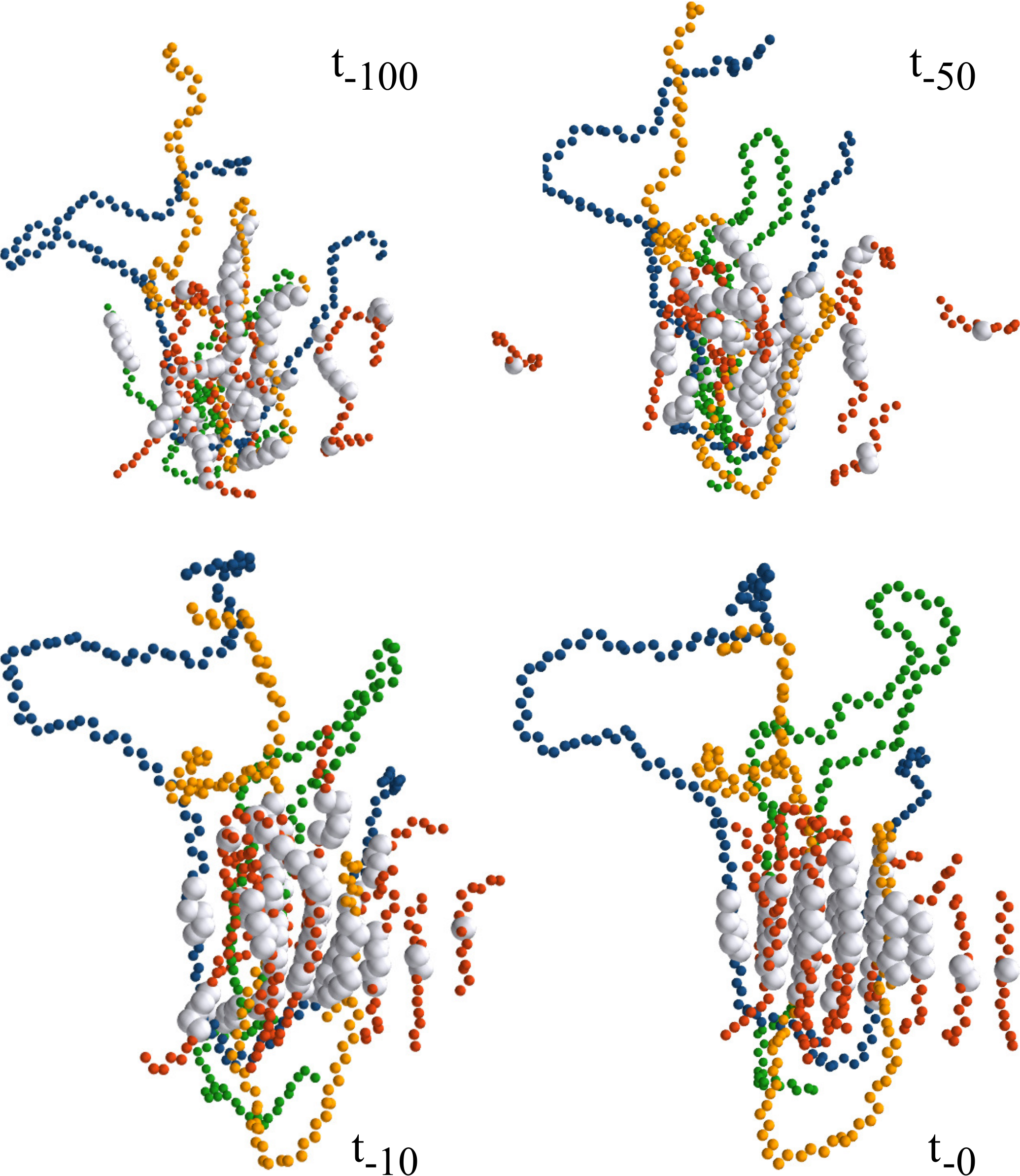}
\caption{Snapshots illustrating the nucleation mechanism. Large gray beads: 
monomers that form the critical nucleus at $t_0$. Red: segments of chains that
participate with a single stem in the formation of the critical nucleus.
Blue, green and orange: chains which fold back and 
participate in the formation of the critical nucleus with more than one stem. 
For the case of folded chains we show complete chains instead of 
segments so that folds and tails can be identified.}
\label{fig:fig_2}
\end{figure}

In Fig.~\ref{fig:fig_2} we present snapshots of the formation of the critical 
nucleus at different times from $t=t_{-100}$ to $t=t_0$. 
The monomers that form the critical nucleus at $t_0$ are highlighted as large 
gray beads. The red color shows the segments of chains that
participate with a single stem in the formation of the critical nucleus
while blue, green and orange indicate those chains which fold back and 
participate in the formation of the critical nucleus with more than one stem. 
For the case of folded chains we show complete chains instead of 
segments so that folds and tails can be identified. The images of the 
formation of the nucleus are consistent with the mechanism we proposed based 
on the values
of $S_2$, $R_{g}$, $V$ and the crystallinity order parameter 
(fig.~\ref{fig:fig_1}).

\begin{figure}[t]
\centering
\includegraphics[width=0.5\textwidth]{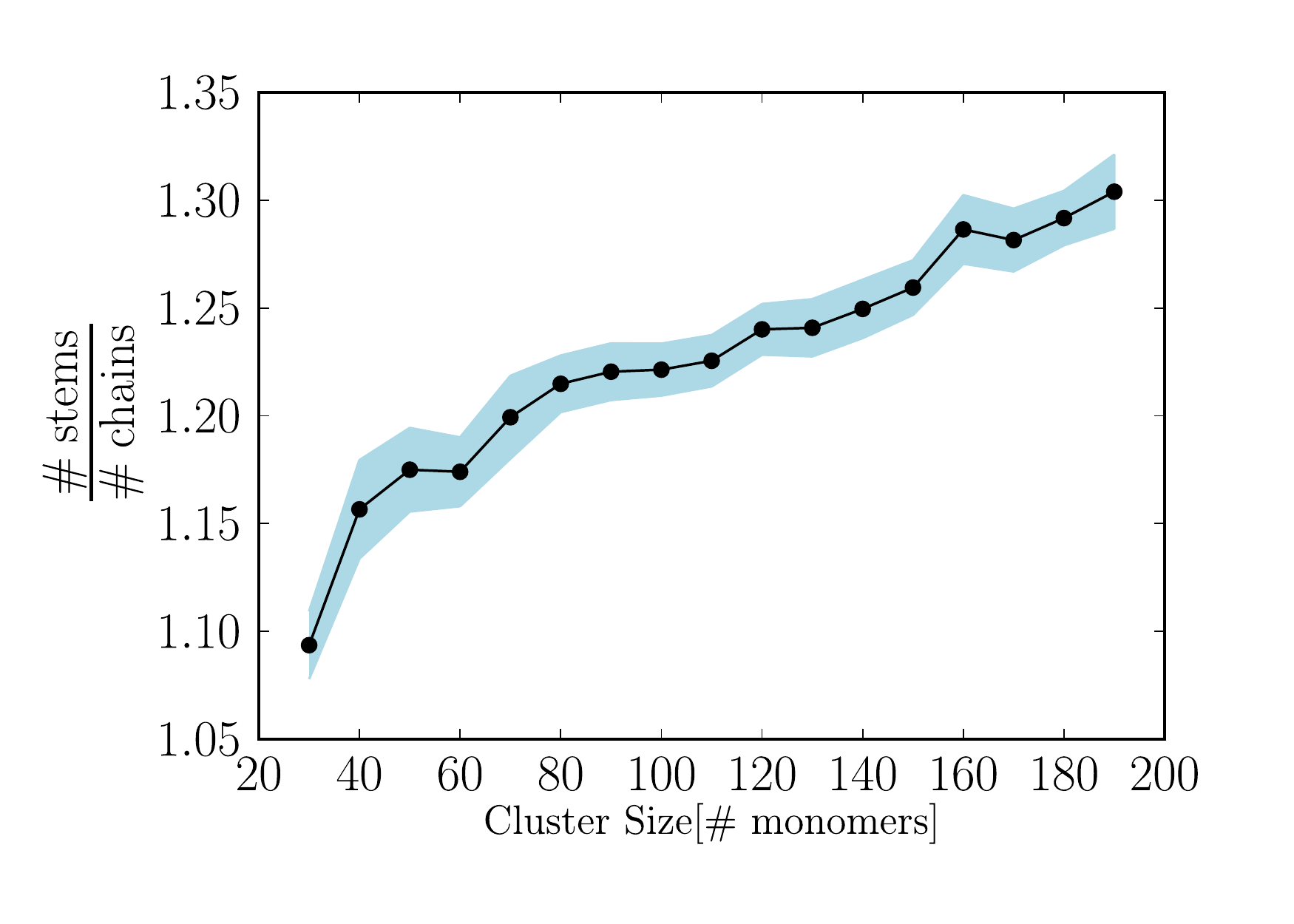}
\caption{Ratio of the number of stems to the number of chains 
against cluster size. The black curve (circles)
shows the mean value and the light blue envelop shows the standard deviation.} 
\label{fig:fig_3}
\end{figure}

The critical nuclei consist of some chain segment (stems) from different chains 
and some from the same chain, which is folded. The primary nucleation mechanism is thus
a combination of intramolecular and intermolecular mechanisms. Fig.~\ref{fig:fig_3} shows the ratio of the number of stems to the number of chains. It is 
always larger than unity, i.e.~there are folded and non-folded chains in the clusters.

\section{C20 under shear}

\subsection{Simulation details}

We studied the effect of shear on the nucleation rate and mechanism in 
$n$-eicosane (C20) 
by means of MD simulations at controlled temperature and constant volume, particle number 
and shear rate in a box with Lees-Edwards boundary conditions. The system 
consisted of 500 chains. We equilibrated it at 450K, which is well above 
the melting temperature. (The equilibrium melting temperature of 
C20 in the simulation model we use is $310\pm2$K \cite{Yi2011}, 
which is in agreement with the experimentally observed value). 
To set the density of the metastable melt at 1 atm pressure we used Table ~\ref{table:table2}. 

We quenched the system from 450K to 250K, applied shear and observed the nucleation event. We ran simulations at seven different shear rates ranging from $\dot{\gamma} = 0.000001\tau^{-1}$ to $\dot{\gamma}=0.01\tau^{-1}$ ($0.95\times10^{10}s^{-1}$ to $0.95\times10^{6}s^{-1}$). 
We also performed simulations at zero shear rate for comparison and we did not find any difference between the nucleation rate at the lowest non-zero shear rate and at zero shear rate.
We used the DPD thermostat \cite{Soddemann2003} with the friction 
coefficient $\gamma_{\rm DPD} = 1.0\tau^{-1}$.  

In order to avoid the artefactual decoupling of the system from its periodic images remarked upon by Chatterjee \cite{Chatterjee2007} when using the DPD thermostat to treat a dissipative shear-flow, a modification to the pairwise dissipative DPD force $\vec{F}^D_{ij}$ was made:

\begin{eqnarray}
\vec{v}_{ij}^{*\alpha}  &=& \vec{v}_{ij}^{\alpha} - \frac{\dot{\gamma}}{L}\vec{r}_{ij}^{\beta}\\
\vec{F}^D_{ij}(\vec{v}_{ij}^{\alpha})  &:=& \vec{F}^D_{ij}(\vec{v}_{ij}^{*\alpha}).
\end{eqnarray}
Where $\vec{v}_{ij}^{*\alpha}$ is laminar flow velocity, $\vec{v}_{ij}^{\alpha}$ is the pairwise velocity parallel to the laminar flow field, $\vec{r}_{ij}^{\beta}$
is the component of pairwise separation perpendicular to the flow field in the shear plane and L is the length of simulation box.
The effect of this modification is to exempt the laminar flow profile from dissipative forces, while allowing dissipation to operate as normal on the flow field with laminar flow subtracted.

\begin{table}
\centering
\caption{\label{table:table2} Density of the metastable melt of $n$-eicosane at 1 atmospheric pressure as a function of temperature.}

\begin{threeparttable}[b]
\begin{tabular}{|c|c|c|}
  \hline
Temperature [$K$] & Density [$g/cm^{3}]$ & Reference\\
\hline
250 & 0.836  & \tnote{*}\\ 
255 & 0.833  & \tnote{**}\\ 
260 & 0.830  & \tnote{**}\\ 
265 & 0.828  & \tnote{*}\\ 
270 & 0.825  & \tnote{**}\\ 
275 & 0.822  & \tnote{**}\\ 
280 & 0.819  & \tnote{*}\\ 

\hline
\end{tabular}
    \begin{tablenotes}
      \item[*] Densities taken from Yi.et .al \cite{Yi2011}
      \item[**] Densities calculated by linear interpolation using data from \cite{Yi2011}. 
    \end{tablenotes}
 \end{threeparttable}

\end{table}

\subsection{Nucleus formation}

\begin{figure}[t]
\centering
\includegraphics[width=0.5\textwidth]{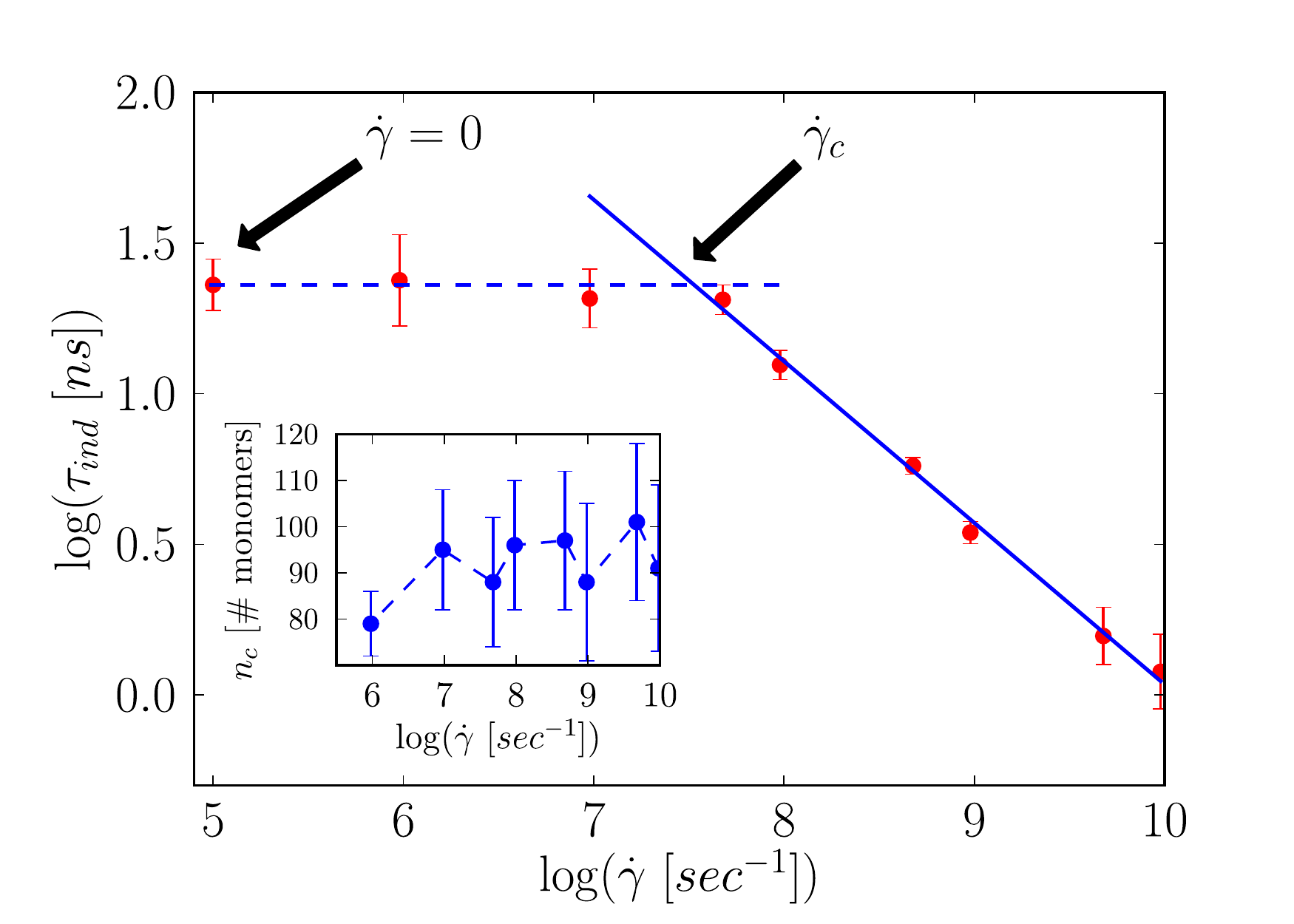}
\caption{Main panel: Induction time versus shear rate for C20 at 250K. Inset: size of critical nucleus versus logarithm of shear rate.}
\label{fig:fig_4}
\end{figure}

Fig.~\ref{fig:fig_4} shows the induction time as a function of shear rate at 
250K. There are two regimes, one in which flow has no effect on the 
induction time, and one where the induction time decreases as a power law 
in the shear rate. This observation agrees with 
experimental results\cite{Coppola2004,Derakhshandeh2012} as well as with the 
theoretical work by Grizzuti and coworkers \cite{Marrucci1983,Coppola2004}. Based on the assumption that shear can only affect nucleation if the sheared chains do not have enough time to relax back into their equilibrium structure, the crossover is expected to occur at Weissenberg number $\tau_{\rm max}\dot{\gamma_c} \approx 1$, where $\tau_{\rm max}$ is the longest relaxation time in the system, and $\dot{\gamma_c}$ is the critical shear rate, at which the induction time begins to drop. In our simulation data $\dot{\gamma}_c$ can be estimated 
from the intersection of the line (continuous) drawn through the induction time data at high shear rates and a horizontal line (dashed) at the value of the induction time under quiescent conditions ($\dot{\gamma}=0$). If we assume that the center of mass diffusion of a chain is the slowest relevant process in the system, we find $\tau_{\rm max}\dot{\gamma_c} = 0.6$, which confirms the assumption. (Here, we have used the time a chain needs to diffuse over the length of its radius of gyration as an estimate of $\tau_{\rm max}$.) 
 
\begin{figure}[t]
\centering
\includegraphics[width=0.5\textwidth]{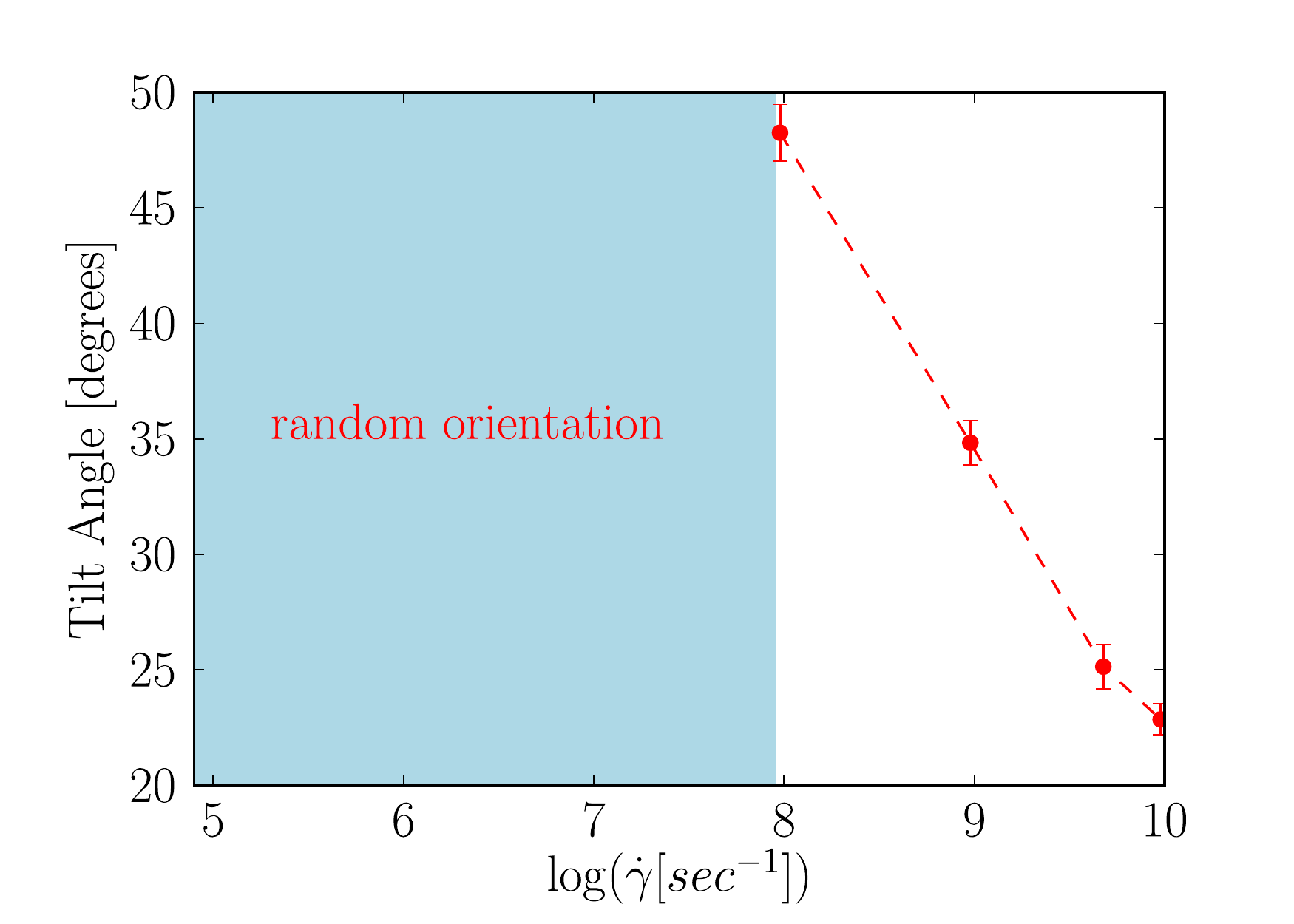}
\caption{C20: Average tilt angle between the critical nucleus and the 
flow direction versus the logarithm of the shear rate. The light blue rectangle shows that the critical nuclei are 
oriented in random directions as in case of quiescent conditions.}
\label{fig:fig_5}
\end{figure}

In agreement with this interpretation, we find that at $\dot{\gamma} < \dot{\gamma}_c$ the nuclei are oriented in any random direction, while at  $\dot{\gamma} > \dot{\gamma}_c$ the nuclei are oriented on average in the direction of flow, i.e.~the stems are parallel to the flow field. 
In Fig.~\ref{fig:fig_5}, we show the average tilt angle of the critical 
nucleus with respect to the flow field at different shear rates.
With increasing shear rate the alignment becomes stronger (this effect is also know from experiments \cite{Manias1995,Peters1995}).
In the inset of Fig.~\ref{fig:fig_4}, the size of the critical nucleus is plotted against the shear rate: there is no effect of shear.


\begin{figure}[t]
\centering
\includegraphics[width=0.5\textwidth]{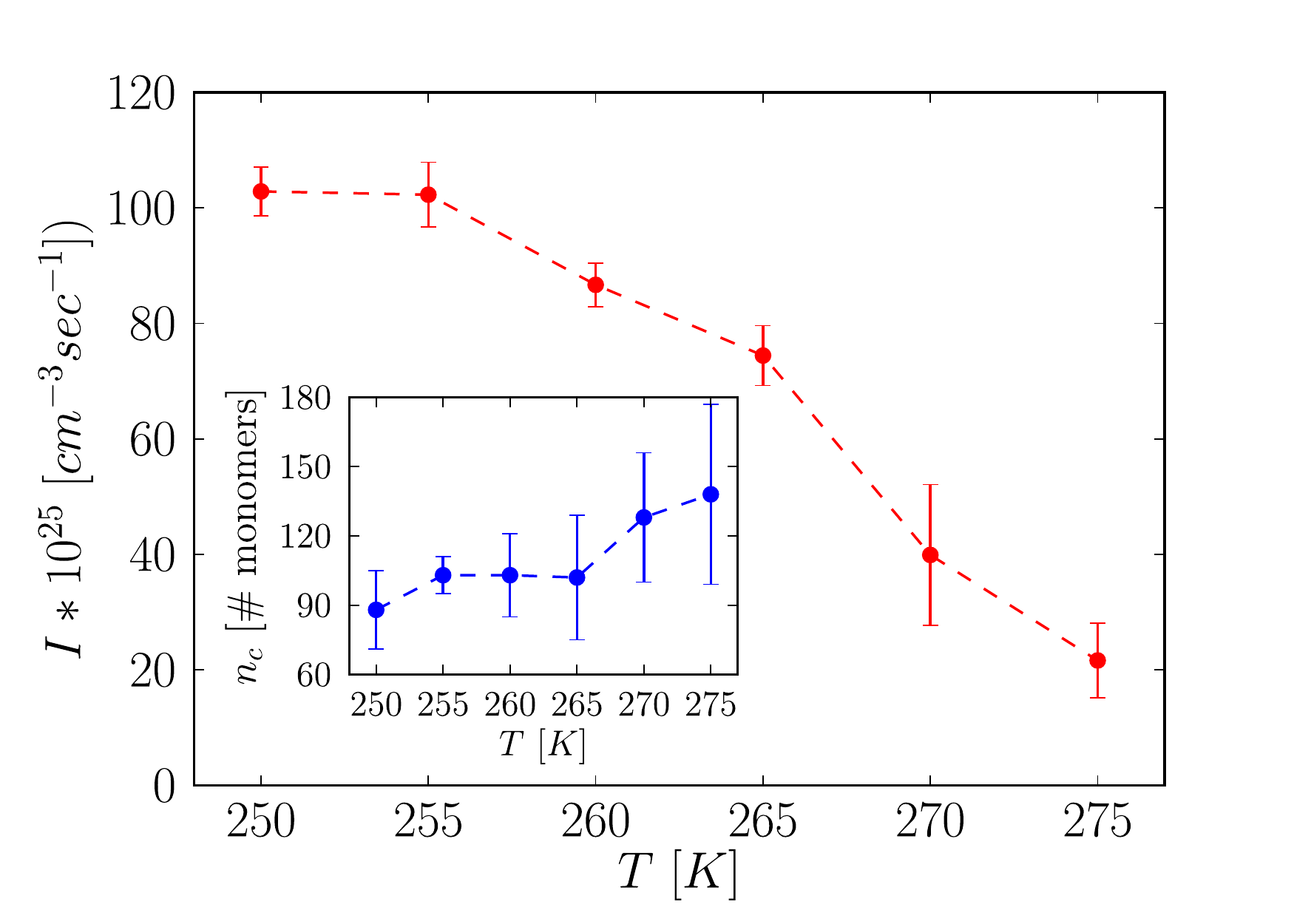}
\caption{Main panel: Nucleation rate versus temperature under shear flow for C20. Inset: size of the critical nucleus versus temperature.}
\label{fig:fig_6}
\end{figure}

Next we discuss the effect of temperature on the nucleation rate under 
shear flow. We carried out simulations at seven different temperatures 
ranging from 250K to 280K 
at a shear rate of $0.001\tau^{-1}(0.95\times10^{9}s^{-1})$. The integration timestep used in 
the simulations at 250K was 0.005$\tau$. 
Fig.~\ref{fig:fig_6} shows the nucleation rate versus temperature. 
As expected, the nucleation rate decreases with increasing temperature. 
However, it increases only by a factor of 5 over the 
temperature range from $250$K to $275$K. As $\gamma=0.001\tau^{-1}$ is well 
above the critical shear rate, the effect of flow on the nucleation 
rate is stronger than the effect of temperature. The chains align primarily 
because they are sheared, and only secondarily because of the chemical potential difference between the bulk crystal and the bulk, metastable melt. (Again this observation is in agreement with experiments and quasi-equilbrium theories \cite{Coppola2004,Derakhshandeh2012}.) 
In the inset of Fig.~\ref{fig:fig_6}, we show the critical nucleus size at 
different degrees of supercooling.  As shear is the dominating driving force for crystallization, the size of the critical nucleus depends only 
weakly on temperature.

\begin{figure}
\centering
\includegraphics[width=0.35\textwidth]{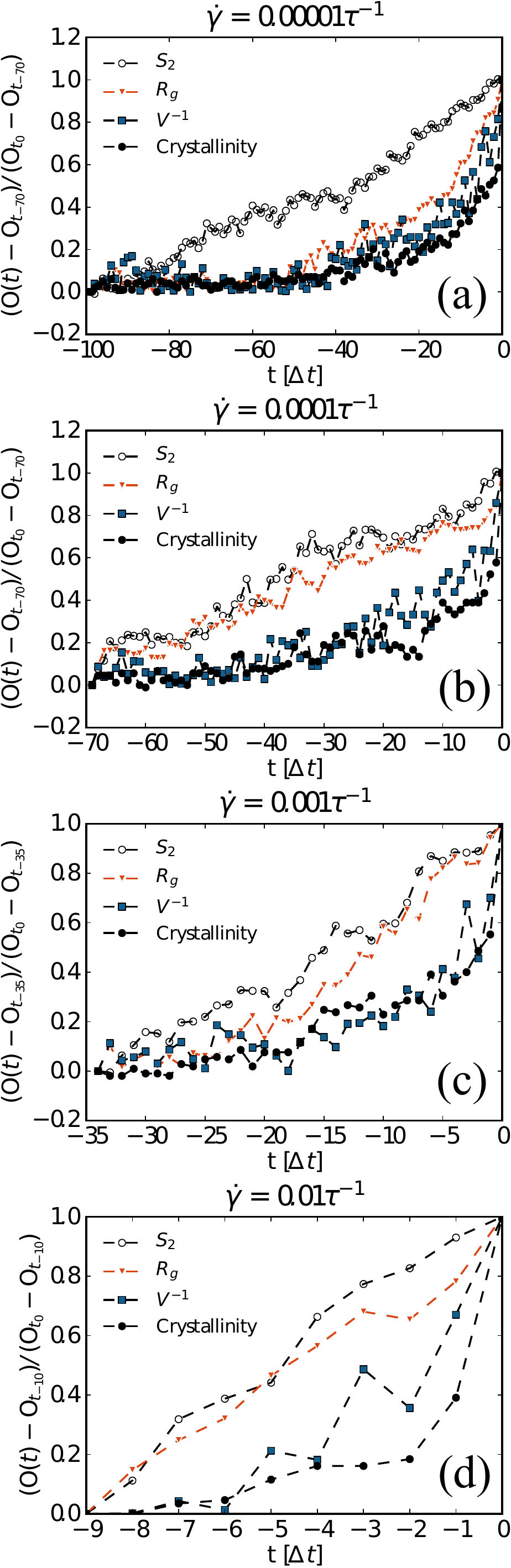}
\caption{C20 under shear: Relative variation of several observables (O) from the melt to the 
formation of a critical nucleus for the particles that are part of the critical nucleus at $t_0$: nematic order $S_{2}$ (black, open circles), the radius of gyration $R_g$ (red, triangles), the inverse of the Voronoi cell volume $V$ (blue, squares) 
and the crystallinity order parameter (black, closed circles). 
(a) : $\dot{\gamma}=0.00001\tau^{-1}$, (b) : $\dot{\gamma}=0.0001\tau^{-1}$, 
(c) : $\dot{\gamma}=0.001\tau^{-1}$, (d) : $\dot{\gamma}=0.01\tau^{-1}$.
The curves are averaged over 10 independent trajectories progressing backwards 
in time from the nucleation time $t=t_{0}$ to $t=-100\Delta t$ at $\dot{\gamma}=0.00001\tau^{-1}$, $t=-70\Delta t$ at $\dot{\gamma}=0.0001\tau^{-1}$, $t=-35\Delta t$ at $\dot{\gamma}=0.001\tau^{-1}$ and $t=-10\Delta t$ at $\dot{\gamma}=0.01\tau^{-1}$ respectively.
Here $\Delta t=10000 \tau$.}
\label{fig:fig_7}
\end{figure}

To study the nucleation mechanism we analyze 10 independent trajectories 
for every shear rate again in terms of the average radius of gyration $R_{g}$ 
of all chains that are part of the nucleus at $t_0$, the nematic order $S_2$ of these chains, the average volume $V$ of the 
Voronoi cell associated to each particle that is part 
of the nucleus and its crystallinity order parameter.
In Fig.~\ref{fig:fig_7} we show the relative variations of these quantities 
with respect to the values they had at $-100\Delta t$, $-70\Delta t$,
 $-35\Delta t$ and  $-10\Delta t$ respectively, where $\Delta t=10000 \tau$, at shear 
rates $\dot{\gamma}=0.00001\tau^{-1}$($0.95\times10^{8}s^{-1}$), 
$\dot{\gamma}=0.0001\tau^{-1}$($0.95\times10^{9}s^{-1}$), 
$\dot{\gamma}=0.001\tau^{-1}$($0.95\times10^{9}s^{-1}$) and 
$\dot{\gamma}=0.01\tau^{-1}$($0.95\times10^{10}s^{-1}$) respectively.
For the lowest shear rate, $\dot{\gamma}=0.00001\tau^{-1}$, on approach to
the formation of the critical nucleus at $t_0$ we observe first an increase 
in the global orientational order $S_2$, then an increase in the radius of 
gyration and in the local density, and finally local positional and 
orientational order are established. 
Thus the nucleation mechanism is the same as in the quiescent 
case \cite{Anwar2013}: first the chains align, then they straighten.

At $\dot{\gamma}=0.0001\tau^{-1}$ and at higher shear rates, we observe a simultaneous  increase in the global orientational order $S_2$ and 
in the radius of gyration $R_{g}$. 
This agrees with the common interpretation of $\dot{\gamma}_c$: when the 
Weissenberg number 
exceeds 1, the chains are straightened and oriented in the 
direction of flow. Thus alignment is enhanced, and the crystallization 
kinetics are accelerated. 

\begin{figure}[t]
\centering
\includegraphics[width=0.5\textwidth]{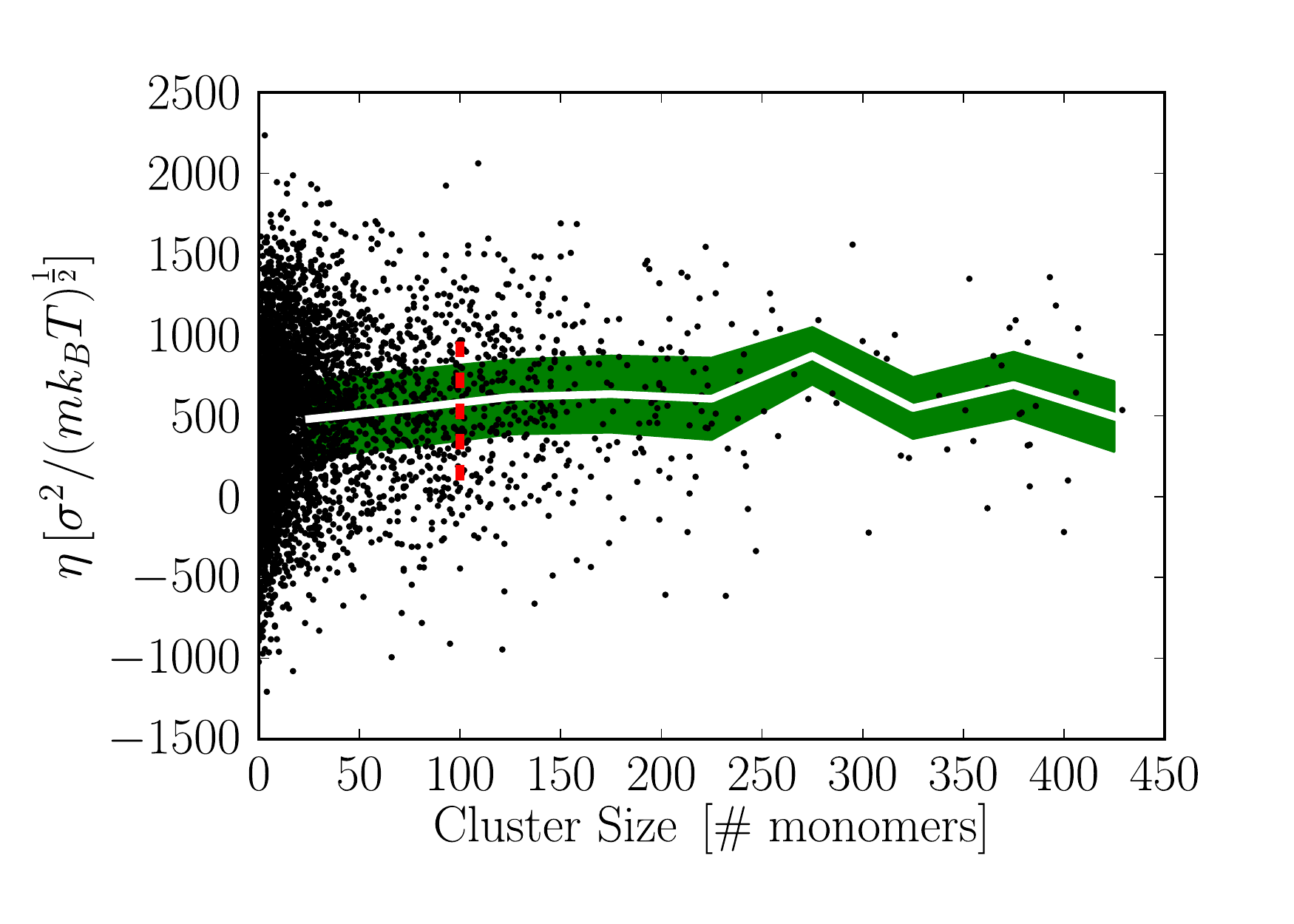}
\caption{C20: Shear viscosity as a function of cluster size. 
Simulation data points (black dots), size of the
critical nucleus (red dashed line), mean value of the viscosity (white line) 
and its standard deviation (green envelop).}
\label{fig:fig_8}
\end{figure}

We have shown that the flow field has an effect on the nucleation rate. In turn, 
the presence of the nucleus should also have an effect on the flow field, because the mechanical properties of a crystal differ considerably from those of the melt.
In fig.~\ref{fig:fig_8} we show the shear viscosity (measured using the instantatneous system average of the stress tensor) 
as a function of cluster size for a system consisting of 500 chains of C20, at 250K and 
at a shear rate of $0.001 \tau^{-1}$. The black dots represent the simulation data points (which are subject to strong fluctuations due to the small system size), the red dashed line shows the size of the critical nucleus, the
white line represents the mean value of the viscosity and the green envelop around the white line represents the standard deviation. 
We do not observe any change in the viscosity during the formation of the nucleus and growth up to a cluster size of 450 monomers. Above this cluster size the
scalar pressure started to decrease, because the phase transition was 
simulated in the NVT ensemble. We conclude that the nucleation events do not 
have an effect on the flow field, as the nuclei are small for the 
temperatures that we discuss here.

\section{C150 under shear}

We performed simulations of 100 chains of C150, equilibrated the system at 500K and then quenched it to 280K. All simulations were carried out under constant volume and temperature conditions at a density of 0.89 $g/cm^3$. We applied shear rates $\dot{\gamma}$  ranging from 0.0001$\tau^{-1}$ to 
0.005$\tau^{-1}$ ($1.012\times 10^{8}s^{-1}$ to $5.06\times 10^{9}s^{-1}$). 

\begin{figure}[t]
\centering
\includegraphics[width=0.5\textwidth]{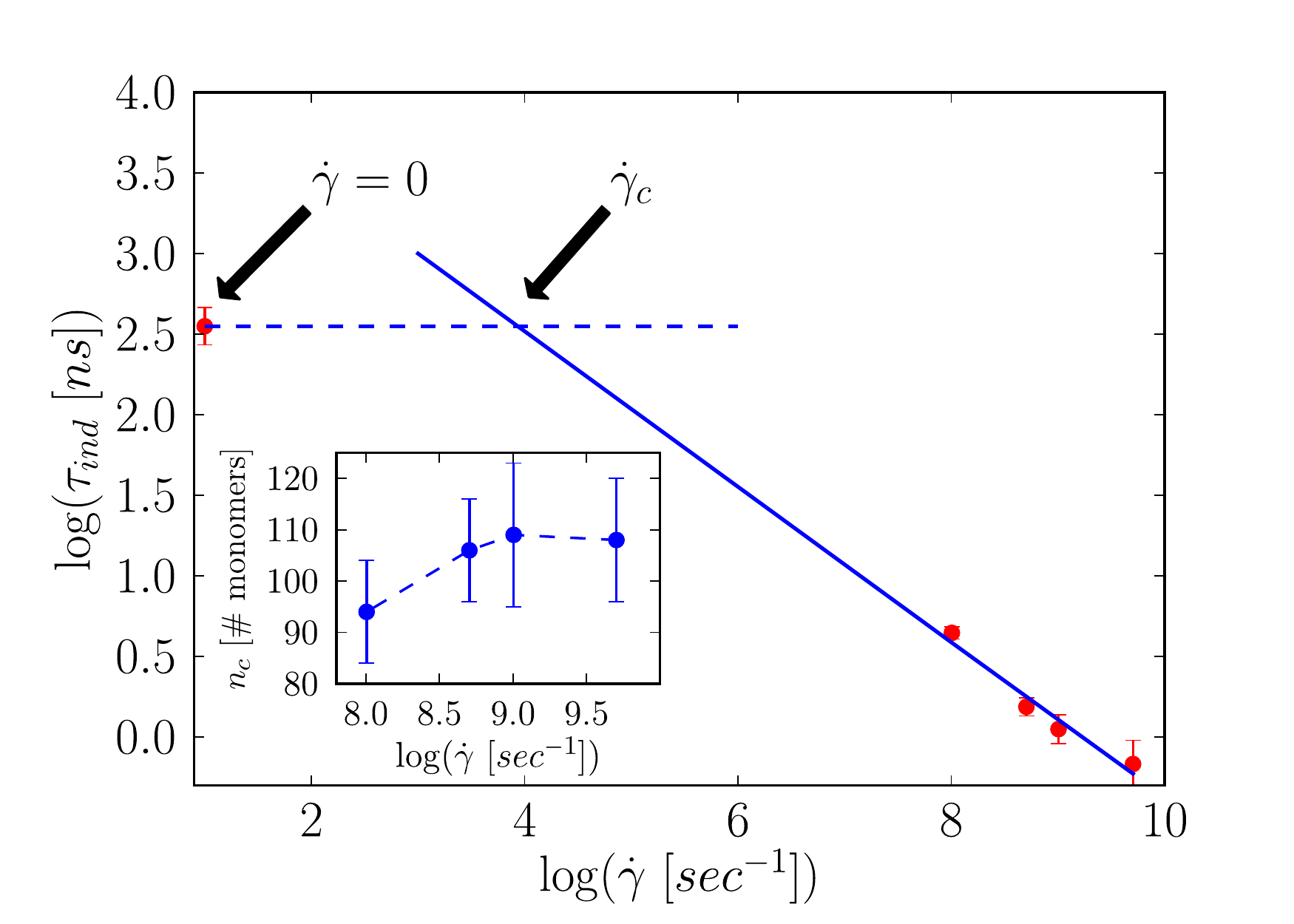}
\caption{Main panel: Induction time against shear rate for C150. Simulation data (red cricles) and fit (blue line). Inset: Size of the critical nucleus against log of shear rate.}
\label{fig:fig_9}
\end{figure}

\begin{figure}[t]
\centering
\includegraphics[width=0.5\textwidth]{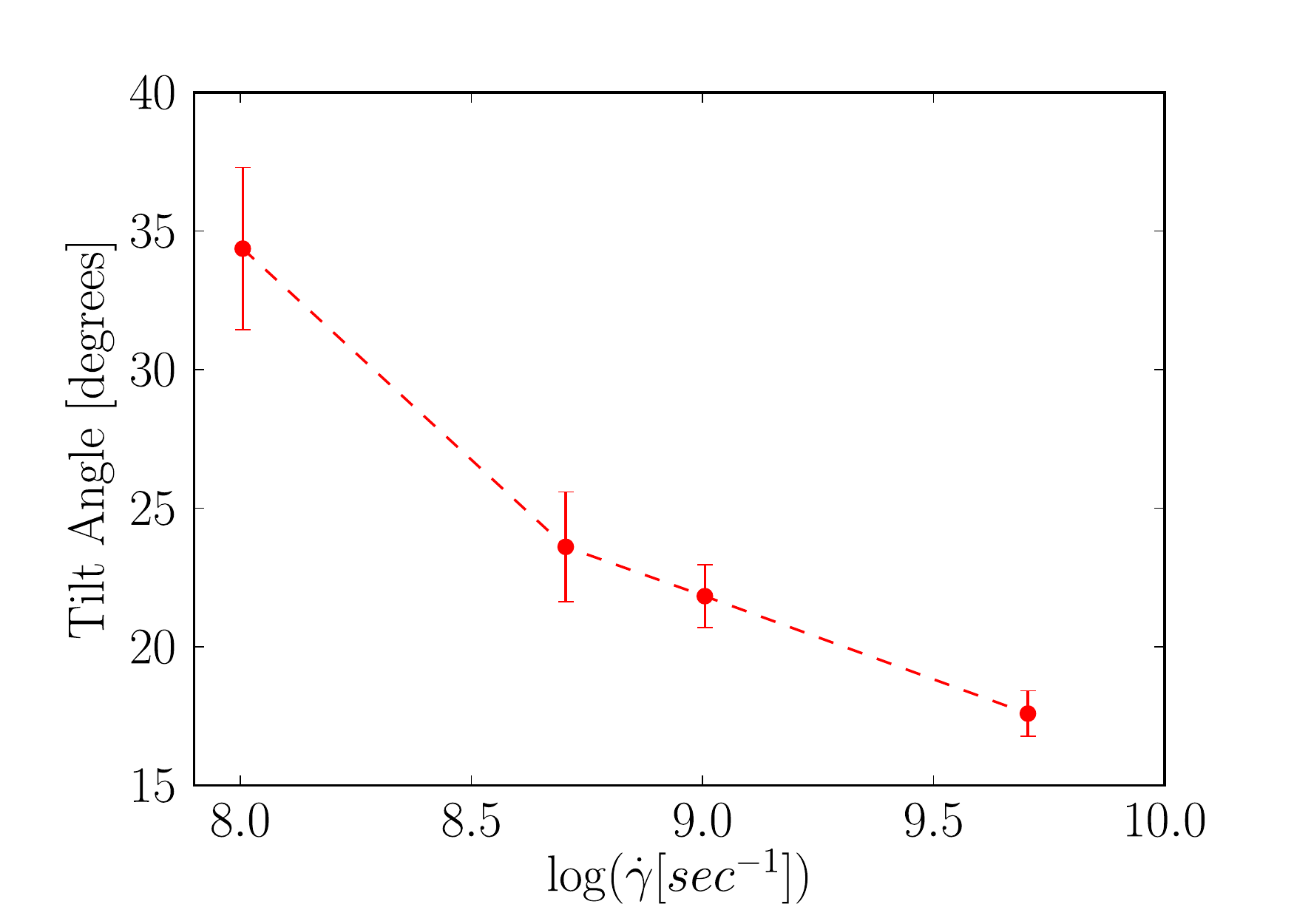}
\caption{C150: Tilt angle of nucleaus versus shear rate.}
\label{fig:fig_11}
\end{figure}

In Fig.~\ref{fig:fig_9} we show the induction time versus the shear rate. The red circles represent the simulation data points and the blue line is a fit. Again, the critical shear rate can be estimated 
as the intersection of the fitted line (continuous) at higher shear rate and a horizontal line (dashed) placed at the value of the induction time under quiescent conditions ($\dot{\gamma}=0$). (The data point at $\dot{\gamma}=0$ is the same as in Table~\ref{tab:table1}.)
In the inset of Fig.~\ref{fig:fig_9}, the size of the critical nucleus is shown versus the shear rate. It is constant within the error bars. And above $\dot{\gamma}_c$, the crystallites are again aligned with the flow field (fig.~\ref{fig:fig_11}). Thus all results are qualitatively the same as those shown in Fig.~\ref{fig:fig_4} for C20. Quanititatively, however, there is a difference: if we take the time the center of mass of a chain needs to diffuse across its radius of gyration to estimate the Weissenberg number at the critical shear rate, we obtain $\tau_{\rm max}\dot{\gamma}_c = 0.007 \ll 1$. We do not interpret this as a contradiction to the theory, but rather as a finite-size effect. The induction time at zero shear rate has been obtained at constant pressure, while the induction times under shear have been obtained at constant volume. The system sizes that we simulated were relatively small, thus even though the average densities and pressures were matched, there were large fluctuations (much larger than in the case of C20, where we could work with more chains). We therefore assume that our estimate of $\dot{\gamma_c}$ is not accurate.

\begin{figure}
\centering
\includegraphics[width=0.35\textwidth]{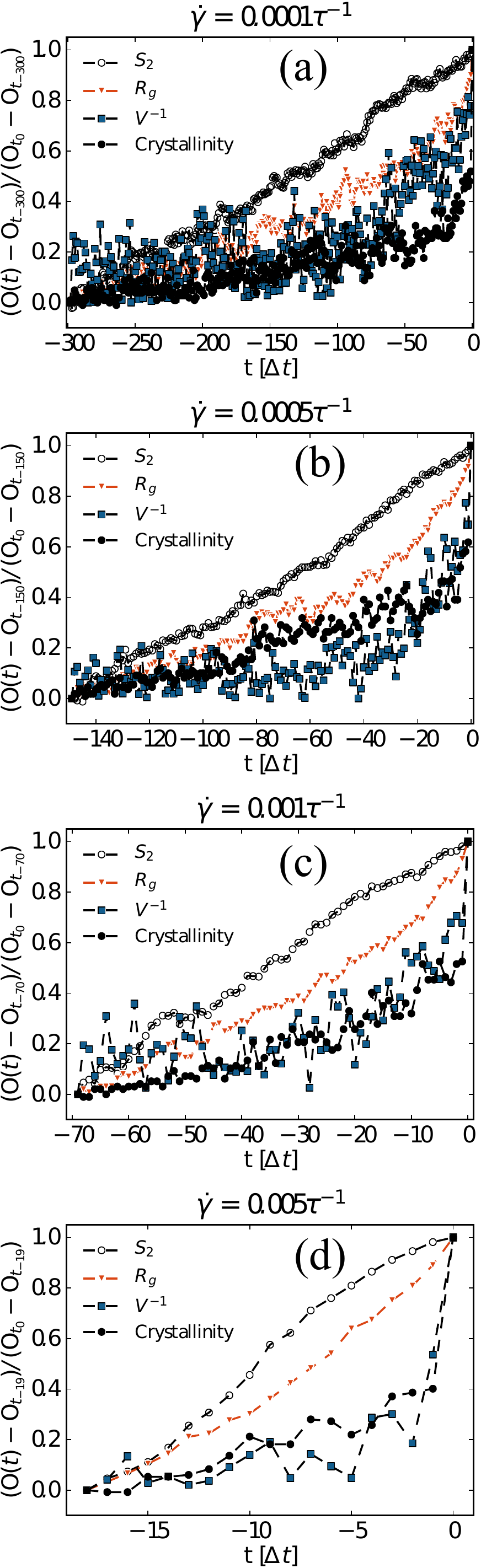}
\caption{C150: Relative variation of several observables (O) from the melt to the formation of a critical nucleus computed for the particles involved 
in the nucleus: orientational order $S_{2}$ (black, open circles), radius of gyration $R_g$ (red, triangles), the inverse of the Voronoi cell volume $V$ (blue, squares) and the crystallinity order parameter (black, close circles). (a) : $\dot{\gamma}=0.0001\tau^{-1}$, (b) : $\dot{\gamma}=0.0005\tau^{-1}$, 
(c) : $\dot{\gamma}=0.001\tau^{-1}$, (d) : $\dot{\gamma}=0.005\tau^{-1}$.
The curves are averaged over 10 independent trajectories progressing backward 
in time from the nucleation time $t=t_{0}$ to $t=-300\Delta t$ at $\dot{\gamma}=0.0001\tau^{-1}$, $t=-140\Delta t$ at $\dot{\gamma}=0.0005\tau^{-1}$, $t=-70\Delta t$ 
at $\dot{\gamma}=0.001\tau^{-1}$ and $t=-20\Delta t$ at $\dot{\gamma}=0.005\tau^{-1}$ respectively. Here $\Delta t=5000 \tau$.}
\label{fig:fig_10}
\end{figure}

To identify the nucleation mechanism, we analyze 10 independent trajectories for every shear 
rate in terms of the order parameters that we introduced in the previous section.
In Fig.~\ref{fig:fig_10} we show the relative variations of these quantities with respect to the values they had at $-300\Delta t$, $-140\Delta t$,
 $-70\Delta t$ and  $-20\Delta t$ respectively, where $\Delta t=5000 \tau$  at shear 
rates $\dot{\gamma}=0.0001\tau^{-1}$($1.012\times 10^{8}s^{-1}$), 
$\dot{\gamma}=0.0005\tau^{-1}$($5.06\times 10^{8}s^{-1}$), 
$\dot{\gamma}=0.001\tau^{-1}$($1.012\times 10^{9}s^{-1}$) and $\dot{\gamma}=0.005\tau^{-1}$($5.06\times 10^{10}s^{-1}$) respectively.
For all shear rates, we observe first an 
increase in the nematic order $S_2$, then an increase in the 
radius of gyration and in the local density, and finally the
crystal structure with local order is formed.

\begin{figure}[t]
\centering
\includegraphics[width=0.5\textwidth]{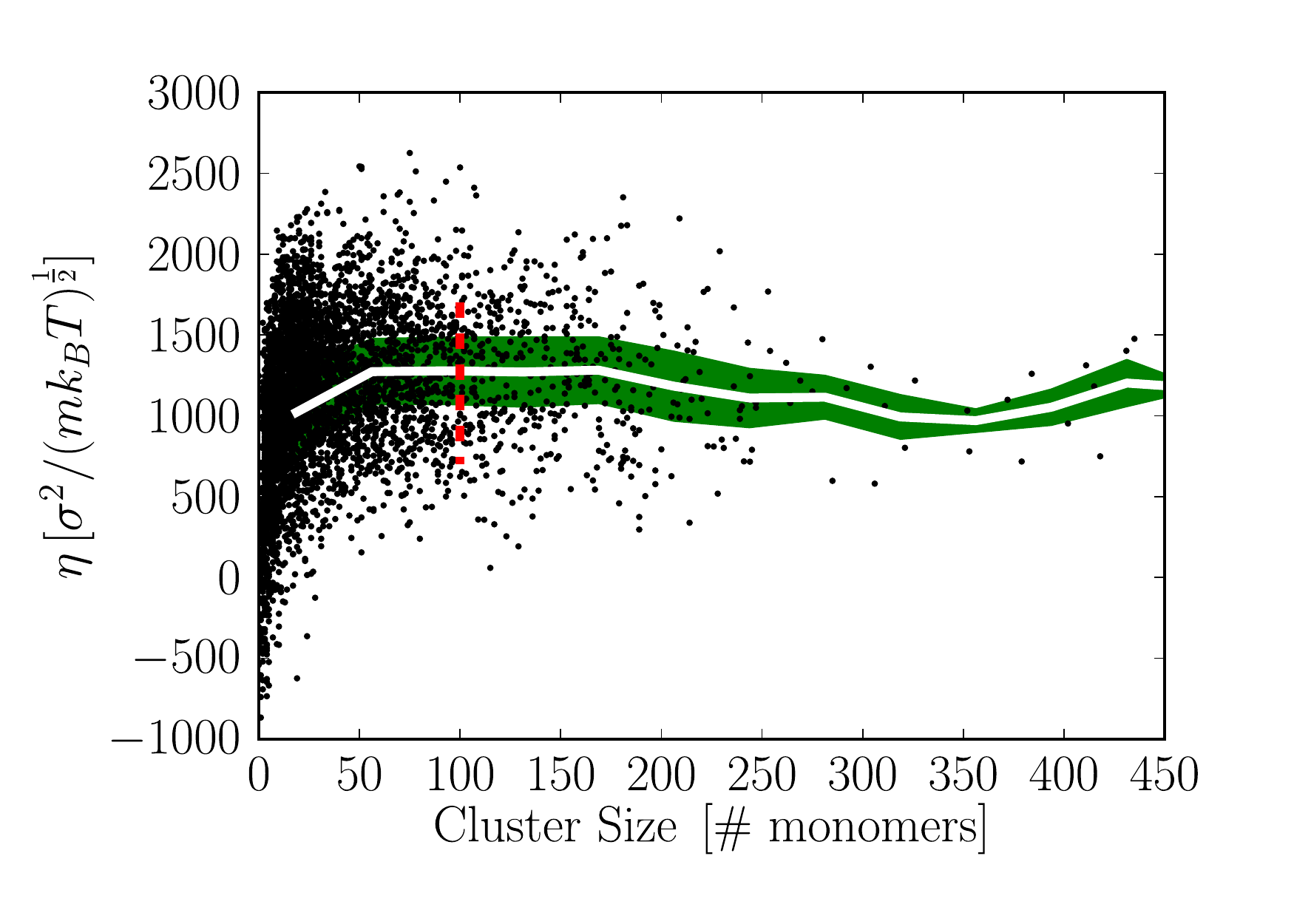}
\caption{C150: Shear viscosity as a function of cluster size. 
Simulation data points (black dots), size of the
critical nucleus (red dashed line), mean value of the viscosity (white line) 
and its standard deviation (green envelop).}
\label{fig:fig_12} 
\end{figure}

To conclude, we show in fig.~\ref{fig:fig_12} the shear viscosity as 
a function of cluster size at a shear rate of $0.001 \tau^{-1}$.  
Again we do not observe any change in the viscosity during the formation of the nucleus and growth up to cluster size of 450 monomers.

\section{Conclusions}
We have simulated crystal nucleation from undercooled melts of short polymer chains under quiescent and shear conditions and analyzed the formation of the critical nucleus. For C150, which is longer than the entanglement length, we observe the same nucleation mechanism as for C20\cite{Anwar2013}, which is shorter than the entanglement length: under quiescent conditions, first the chain segments align, then they straighten, and finally the cluster becomes denser and local positional and orientational order are established.

At low shear rates we observe the same nucleation mechanism as under quiescent conditions while at high shear rates the chains (or chain segments) 
align and straighten at the same time, then the local density increases and finally local positional and orientational order are established. We estimate the
critical shear rates for both systems(C20 \& C150) and find power law behaviour between nucleation rate and shear rate in agreement with experiments and theory\cite{Coppola2004}.

\begin{acknowledgments}
We thank Francesco Turci, Jens-Uwe Sommer, Roland Sanctuary, J\"org Baller and 
Carlo Di Giambattista for stimulating discussions. 
This project has been financially supported by the National Research Fund (FNR) within the CORE project Polyshear. 
Computer simulations presented in this paper were carried out using the HPC facility of the University of Luxembourg.
\end{acknowledgments}
%Merlin.mbs v4.21 2009-07-09.

%\bibliographystyle{myaps}
%\bibliography{mybiblio}

%

\end{document}